\documentclass[aps, prx, twocolumn, a4paper, 10pt, showpacs,reprint,superscriptaddress,nofootinbib]{revtex4-2}

\usepackage[left=2cm, right=2cm, top=2.5cm, bottom=2.5cm]{geometry}

\usepackage{amsmath}
\usepackage{amssymb}
\usepackage{mathtools}
\usepackage{physics}
\usepackage{blkarray}

\usepackage{graphicx}
\usepackage{subcaption}
\usepackage[dvipsnames]{xcolor}
\usepackage{tikz}
\usetikzlibrary{matrix,positioning}

\usepackage{listings}
\usepackage{enumerate}

\usepackage{hyperref}

\begin{document}

\title{A QSVT-Based Quantum Jacobi Algorithm for Linear Systems with Application to the Poisson Equation}

\author{Louisa M. Piskol}
\email{louisa.marie.piskol@volkswagen.de}
\affiliation{Volkswagen AG, Berliner Ring 2, 38440 Wolfsburg, Germany}
\affiliation{Institute for Partial Differential Equations, Technical University Braunschweig, Universitätsplatz 2, 38106 Braunschweig, Germany}

\author{Thorsten Grahs}
\affiliation{Volkswagen AG, Berliner Ring 2, 38440 Wolfsburg, Germany}

\author{Stefan Langer}
\affiliation{Institute for Aerodynamics and Flow Technology, German Aerospace Center (DLR), Lilienthalplatz 7, 38108 Braunschweig, Germany}

\author{Oleksandr Kyriienko}
\affiliation{School of Mathematical and Physical Sciences, University of Sheffield, Sheffield S3 7RH, United Kingdom}

\date{\today}

\begin{abstract}
Many computational fluid dynamics (CFD) algorithms solve partial differential equations by discretization, resulting in large and sparse systems of linear equations. While iterative methods are widely used to solve these systems classically, most existing quantum linear system solvers target the solution through matrix inversion rather than approximating it using an iterative procedure. In this work, we develop a quantum implementation of the Jacobi method based on quantum singular value transformation (QSVT). By reformulating the Jacobi iteration as a polynomial transformation of a block-encoded operator, the algorithm requires only a constant ancilla overhead with respect to the number of iterations while maintaining a circuit depth that scales linearly with the iteration number. We demonstrate the algorithm for one- and two-dimensional Poisson problems, including the pressure Poisson equation arising in Chorin's projection method for the lid-driven cavity flow. The proposed algorithm provides a promising building block for future quantum implementations of multigrid methods and preconditioning techniques, bringing quantum algorithms closer to established CFD solution strategies.
\end{abstract}

\maketitle
\pagenumbering{arabic}

\section{Introduction}

The ability to accurately predict fluid flows is essential in many areas of science and engineering, ranging from aerodynamic design to environmental engineering and energy applications \cite{anderson_computational_1995}. The underlying fluid dynamics are commonly described by systems of partial differential equations (PDEs), such as the Navier-Stokes equations \cite{ferziger_computational_2020}. In computational fluid dynamics (CFD), these equations are solved numerically to simulate fluid flow. This reduces the need for costly real-world experiments and accelerates engineering design, making CFD an important tool for analysing and predicting complex flow phenomena \cite{anderson_computational_1995}.
However, accurately predicting fluid dynamics remains challenging due to the nonlinear and multiscale nature of the governing equations. In particular, turbulent flows and boundary-layer interactions require fine spatial and temporal resolution \cite{wilcox_turbulence_2010}. As a result, realistic CFD simulations quickly become computationally expensive.

Most CFD approaches discretize the computational domain into a grid using finite difference, finite volume, or finite element methods \cite{ferziger_computational_2020}. This transforms the governing PDEs into large and sparse systems of linear equations. For large-scale CFD simulations, these systems are typically solved by using iterative methods rather than direct methods, often in combination with suitable preconditioners \cite{saad_iterative_2003, benzi_preconditioning_2002}. In addition, multigrid methods are widely used to further accelerate convergence \cite{wesseling_introduction_1992}. In this context, simple stationary iterations, such as the Jacobi and Gauss–Seidel methods, remain important and are commonly employed as smoothers \cite{wesseling_geometric_2001}.

Quantum computing represents a promising paradigm for accelerating scientific simulations that are challenging for classical computers \cite{nielsen_quantum_2010}. The interest in quantum algorithms for CFD has been increasing in recent years, rooted in their ability to encode and propagate coherently fine-grid problems in the exponentially large Hilbert space of quantum systems \cite{succi_quantum_2023, bharadwaj_simulating_2025}. 
Several approaches focus on solving the governing differential equations directly, including quantum time-marching methods \cite{fang_time-marching_2023}, algorithms based on the linear combination of Hamiltonian simulations (LCHS) for linear non-unitary differential equations \cite{an_linear_2023, an_quantum_2026}, and Schrödingerisation, which reformulates the differential equations as Schrödinger equations \cite{jin_quantum_2023, hu_quantum_2024}.
Another approach is the quantum lattice Boltzmann method, where fluid dynamics are described in terms of discrete particle distribution functions following its classical counterpart, and where the streaming and collision steps are implemented using quantum circuits \cite{mezzacapo_quantum_2015, li_potential_2025}. 

Inspired by the growing use of machine learning in CFD \cite{vinuesa_enhancing_2022, cuomo_scientific_2022}, variational quantum algorithms have also been proposed for solving differential equations. These methods employ parametrized quantum circuits, whose parameters are optimised to approximate the solution of the governing equations \cite{lubasch_variational_2020, jaksch_variational_2023}. A related direction is given by quantum physics-informed neural networks (QPINNs), which incorporate the differential equation and boundary conditions directly into the loss function \cite{kyriienko_solving_2021, paine_quantum_2023, paine_physics-informed_2023, siegl_solving_2025, berger_trainable_2025}. 

Another distinct class of quantum algorithms for CFD-type are quantum linear system solvers (QLSSs) \cite{morales_quantum_2025}, which can be applied to the large and sparse linear systems arising from the discretization of partial differential equations. Given a linear system
\begin{equation}
    Ax=b
\end{equation}
with $A\in\mathbb{C}^{N\times N}$ and $x,b\in\mathbb{C}^{N}$, the goal of a QLSS is to prepare a quantum state $\ket{x}$ proportional to the solution vector $x$. This problem was first addressed by the HHL algorithm \cite{harrow_quantum_2009}. Since then, numerous improvements have been proposed \cite{ambainis_variable_2010, childs_quantum_2017, gilyen_quantum_2019}. These include direct inversion methods based on linear combinations of unitaries \cite{childs_quantum_2017} and quantum singular value transformation (QSVT) \cite{gilyen_quantum_2019}, as well as algorithms based on adiabatic quantum computing (AQC) \cite{subasi_quantum_2019}. Recent adiabatic \cite{costa_optimal_2022, jennings_randomized_2025} and adiabatic-inspired \cite{dalzell_shortcut_2024} approaches have achieved the optimal complexity in terms of condition number and target precision, matching the known theoretical lower bound for solving quantum linear systems. 

Given the central role of iterative methods in modern CFD, extending the quantum algorithmic toolbox beyond inversion methods is of particular interest. Initial work considered stationary iterative methods through Schrödingerisation and Hamiltonian simulation \cite{jin_quantum_2024}. More recently, Williams \textit{et al.} \cite{williams_quantum_2026} presented the first gate-based quantum implementation of simple iterative methods, focusing on the Jacobi method, using block encodings and the linear combination of unitaries (LCU). However, in this construction, the Jacobi iterations are implemented through products of block encodings. Since each block encoding introduces its own ancilla register, the number of ancilla qubits grows linearly with the number of iterations. 

In this work, we propose an alternative implementation of a quantum Jacobi algorithm based on QSVT, which efficiently implements polynomial transformations of the singular values of block-encoded operators. By reformulating the iteration as a polynomial transformation of a block-encoded operator, the required number of ancilla qubits becomes constant with respect to the number of Jacobi iterations, while also reducing the circuit depth. 

After introducing the necessary background on block encodings, QSVT, and the previous quantum Jacobi algorithm (Sec. \ref{sec:background}), we derive the proposed algorithm from the classical Jacobi iteration and present its quantum circuit implementation (Sec. \ref{sec:QJacobi}). We then analyse its quantum resource requirements and compare them with the existing approach. The algorithm is demonstrated for simple one-dimensional Poisson problems and, as a more physically relevant example in CFD, for the two-dimensional Poisson equation arising in Chorin's projection method for the lid-driven cavity problem (Sec. \ref{sec:Application}). Finally, we discuss how the polynomial-based quantum implementation can be extended to other iterative methods (Sec. \ref{sec:Discussion}).

\section{Preliminaries and Background}\label{sec:background}

To establish the necessary background, we briefly introduce the fundamentals of block encodings and quantum singular value transformation, which form the basis of the proposed algorithm. Additionally, we summarize the classical Jacobi method together with previous work on its quantum implementation.
\begin{figure*}
    \centering
    \includegraphics[width=0.8\linewidth]{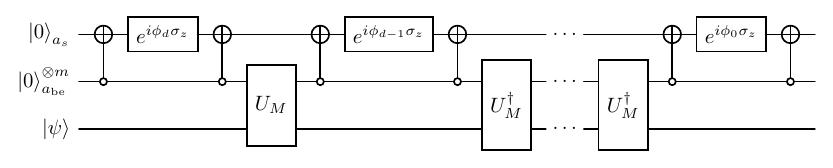}
    \caption{Circuit representation of the Quantum Singular Value Transformation (QSVT) for constructing a block encoding of $\textup{Poly}(M)$. The projector-controlled phase-shift operators are implemented using the signal qubit $a_s$, two controlled-NOT operations, and a single-qubit phase gate $e^{i\phi\sigma_z}$. Here $d$ is even. When $d$ is odd, the last $U_M^\dag$ gate should be replaced with a $U_M$ gate.}
    \label{fig:QETcircuit}
\end{figure*}

\subsection{Block encodings}\label{sec:BlockEncoding}

Block encoding is a fundamental technique in quantum computing that enables the implementation of general, non-unitary matrices on a quantum computer \cite{berry_black-box_2012, low_hamiltonian_2019, gilyen_quantum_2019}. Since quantum circuits are restricted to unitary operations, a given matrix must first be embedded into a larger unitary operator acting on an extended Hilbert space. This concept forms the basis of many modern quantum algorithms, including the quantum singular value transformation (QSVT) framework that will be discussed in the following subsection.

In general, for any $n$-qubit matrix $M$ satisfying $\norm{M}<1$ we can construct a unitary operator through the following unitary dilation by using an ancillary qubit,
\begin{equation}\label{eq:BlockEncoding}
U_M =
\begin{pmatrix}
M & \sqrt{I - M M^\dagger} \\
\sqrt{I - M^\dagger M} & -M^\dagger
\end{pmatrix}.
\end{equation}
However, this dilation is typically not suitable for efficient implementation in a quantum circuit. More generally, a unitary $U_M$ using $m$ ancilla qubits is called a $(\alpha, m, \varepsilon)$-block encoding of $M$ if
\begin{equation}
    \norm{M - \alpha(\bra{0}^{\otimes m}\otimes I)U_M(\ket{0}^{\otimes m}\otimes I)}\leq\varepsilon,
\end{equation}
where $\alpha$ is a subnormalization factor and $\varepsilon$ is an approximation error \cite{gilyen_quantum_2019}.

There are various techniques to construct block encodings \cite{childs_hamiltonian_2012, camps_fable_2022, camps_explicit_2024, sunderhauf_block-encoding_2024}. A widely used approach is to use the linear combination of unitaries (LCU), where the matrix is represented as a weighted sum of unitary operators \cite{childs_hamiltonian_2012}. Other approaches take advantage of structural properties of the matrix, such as sparsity \cite{camps_explicit_2024, sunderhauf_block-encoding_2024}. In the context of computational fluid dynamics, specialized block encodings have also been developed for discretized differential operators such as the Laplacian, and several implementation strategies have been compared in the literature \cite{ sunderhauf_block-encoding_2024, lapworth_evaluation_2024, sturm_efficient_2025, boutot_explicit_2026}.

\subsection{Quantum Singular Value Transformation}\label{sec:QSVT}

Quantum singular value transformation (QSVT) is a framework for implementing polynomial transformations of the singular values of a matrix \cite{gilyen_quantum_2019, martyn_grand_2021}.
It offers a unified approach to a variety of quantum algorithms, such as Hamiltonian simulation and matrix inversion, the latter of which can be used to solve linear systems of equations.

Given a matrix in its singular value decomposition $M=\sum_k \sigma_k \op{w_k}{v_k}$, QSVT constructs a polynomial transformation by alternating applications of the block encoding $U_M$ and the projector-controlled phase-shift operators $\Pi_\phi$ and $\widetilde{\Pi}_\phi$.
The resulting sequence implements a polynomial of degree at most $d$, provided that the target polynomial has a definite parity (either even or odd). The phase angles $\vec{\phi}\in\mathbb{R}^{d+1}$ are chosen such that the desired polynomial is realized and can be computed using established classical preprocessing methods \cite{chao_finding_2020, martyn_grand_2021, dong_robust_2024}. 

More precisely, for even $d$, the QSVT sequence is defined as
\begin{equation}
    U_{\Vec{\phi}} = \Pi_{\phi_{0}} \left[ \prod_{j=1}^{d/2} U_M^\dag \Tilde{\Pi}_{\phi_{2j-1}} U_M \Pi_{\phi_{2j}} \right],
\end{equation}
while for odd $d$ it is given by
\begin{equation}
    U_{\Vec{\phi}} = \Tilde{\Pi}_{\phi_{0}} U_M \Pi_{\phi_{1}} \left[ \prod_{j=1}^{(d-1)/2} U_M^\dag \Tilde{\Pi}_{\phi_{2j}} U_M \Pi_{\phi_{2j+1}} \right].
\end{equation}
The resulting operator is a block encoding of
\begin{equation}
    \textup{Poly}(M)=
    \begin{cases}
        \sum_k \textup{Poly}(\sigma_k) \dyad{v_k},  \quad\text{for even}\, d, \\
        \sum_k \textup{Poly}(\sigma_k) \op{w_k}{v_k},  \quad\text{for odd}\, d,
    \end{cases}
\end{equation}
where $\textup{Poly}(\cdot)$ denotes a polynomial transformation defined on singular values of the operator $M$. 
The corresponding circuit implementation is shown in Figure \ref{fig:QETcircuit}.

In the special case of a Hermitian matrix, the singular-value transformation reduces to an eigenvalue transformation and therefore realizes matrix polynomials in the usual sense, i.e., $\textup{Poly}(M) = \sum_k \textup{Poly}(\lambda_k)\dyad{v_k}$, where $\lambda_k$ are the eigenvalues of $M$. This property forms the basis of the quantum Jacobi implementation presented in Section \ref{sec:QJacobi}. Additional details on QSVT, its origin from quantum signal processing (QSP) \cite{low_hamiltonian_2019}, the polynomial constraints, and the relation between singular-value and eigenvalue transformations are provided in the Appendix \ref{sec:QSVT_Details}.

\subsection{Jacobi Algorithm}\label{sec:JacobiIntro}

Next, let us provide a background for performing Jacobi iterations. The corresponding algorithm is a stationary iterative splitting method for approximating the solution of a linear system \cite{saad_iterative_2003}. For the linear system $Ax=b$, first the matrix is split into the diagonal matrix $D$ and it`s off-diagonal part $R$, according to $A=D+R$. The solution is then obtained iteratively, with a starting guess $x_0$. The approximation after $k$ iterations is given by
\begin{equation}\label{eq:JacIterativeScheme}
    x_k = D^{-1} (b-R x_{k-1}).
\end{equation}
The iteration converges when the spectral radius of the iteration matrix $M=D^{-1}R$ satisfies $\rho(M)<1$, which is, for example, guaranteed if $A$ is strictly diagonally dominant \cite{saad_iterative_2003}.

The classical Jacobi method can be directly mapped to a quantum version via block encoding, as was shown in Ref. \cite{williams_quantum_2026}. 
The construction requires the state-preparation operators $U_0$ and $U_b$, satisfying $\ket{x_0}=U_0 \ket{0}$ and $\ket{b}=U_b \ket{0}$, as well as the block encodings $U_{D^{-1}}$ and $U_R$ of $D^{-1}$ and $R$, respectively. The inverse of the diagonal matrix $D$ can be computed classically beforehand.
The corresponding block encodings are constructed as described in Sec.~\ref{sec:BlockEncoding} using $m_{D^{-1}}$ and $m_R$ ancilla qubits.

We can write the quantum version for Jacobi iteration scheme as
\begin{equation}
    \ket{x_k} = U_{D^{-1}} \left[ U_b \ket{0} - U_R \ket{x_{k-1}}\right].
\end{equation}
In the classical case one can perform each iteration step after another. However, this becomes impractical for a quantum algorithm, as the read out of a generic state after every iteration is computationally expensive (requires many shots, unless specific properties are measured). Instead, it is beneficial to directly calculate the solution after $k$ iterations, which is given by
\begin{equation}\label{eq:OriginalJacobi}
    \begin{split}
        \ket{x_k} =& \left[ \sum_{j=1}^k \left( -U_{D^{-1}} U_R \right)^{j-1}U_{D^{-1}}U_b \right. \\
        &+  \left. \left( -U_{D^{-1}} U_R \right)^{k} U_0 \right] \ket{0}.
    \end{split}
\end{equation}
This is a sum over $k+1$ unitary terms, e.g. $\sum_{l=1}^{k+1} c_l U_l$, which can be implemented by a linear combination of unitaries and where we can include the normalization factors $\hat{b}=\norm{b}$, $\hat{x}_0=\norm{x_0}$, $\hat{r}=\norm{R}$ and $\hat{d}=\norm{D^{-1}}$ into the LCU coefficients.
In the end, the resulting state $\ket{x_k}$ has to be scaled by the normalization factor $\sum_l c_l$ to recover the correct solution.

To perform the LCU the approach requires $\lceil \log_2(k+1) \rceil$ ancilla qubits. Furthermore, the multiplication of block-encoded matrices requires the ancilla registers of the corresponding block encodings to be treated separately, causing the ancilla requirements to accumulate with every multiplication. As $\left( -U_{D^{-1}} U_R \right)^{k}$ is the largest part in terms of multiplications in Eq.~\eqref{eq:OriginalJacobi}, the multiplication of the block encodings require $k(m_{D^{-1}}+m_R)$ ancilla qubits. Thus, in total, the quantum Jacobi algorithm needs $\log_2(N)+k(m_{D^{-1}}+m_R)+ \lceil \log_2(k+1) \rceil$ qubits, scaling linearly with the number of iterations.

In this formulation, the biggest challenge for the quantum Jacobi algorithm (QJA) becomes the number of iterations stored coherently in the ancillary register of LCU. As the number of iterations grows, this increases the register on which PREPARE operation acts, and probability of post-selecting on correct outcomes of the ancillary register decreases. In the following, we explain how this challenge can be overcome, and suggest an efficient method for QJA-type algorithms.

\section{QSVT-Enhanced Quantum Jacobi Method}\label{sec:QJacobi}

We propose to improve the quantum Jacobi algorithm by replacing the qubit expensive multiplication of block encodings by efficient QSVT subroutines. 
Let us start again from the classical Jacobi scheme of Eq. \eqref{eq:JacIterativeScheme}. By introducing $M = D^{-1} R$ and $\tilde{b} = D^{-1} b$ the scheme becomes
\begin{equation}\label{eq:ClassicJacRewritten}
    \begin{aligned}
        x_k &= \tilde{b} - M x_{k-1} \\
        &= \sum_{j=1}^k (-M)^{j-1} \tilde{b} + (-M)^k x_0.
    \end{aligned}
\end{equation}
The matrix power is equivalent to taking the power of the eigenvalues in the eigendecomposition, e.g. $M^k=V\Lambda^k V^{-1}$. When the matrix $M$ is normal or Hermitian, we can achieve powers of the eigenvalues via a polynomial transformation of the eigenvalues within the QSVT framework. We have already seen that these kind of polynomials must be either even or odd. Therefore, we split Eq. \eqref{eq:ClassicJacRewritten} into three separate terms, one for the even exponents, one for the odd exponents and the last term that acts on $x_0$, which may be either even or odd, depending on $k$:
\begin{equation}\label{eq:JacobiAsPolys}
    \begin{split}
        x_k =& \left( \sum_{j=0}^{\lceil\frac{k}{2}\rceil-1} M^{2j} \right) \tilde{b} - \left( \sum_{j=0}^{\lfloor\frac{k}{2}\rfloor-1} M^{2j+1} \right) \tilde{b} \\
        &+ (-M)^k x_0
    \end{split}
\end{equation}
Now we can construct the quantum solution after $k$ iterations as a combination of three QSVT subroutines as
\begin{equation}\label{eq:FinalJacobi}
    \begin{split}
        \ket{x_k} =& \left[ c_e U_{\mathrm{QSVT}}^e(M) U_{\tilde{b}} + c_o U_{\mathrm{QSVT}}^o(M) U_{\tilde{b} } \right. \\
        & \left. + c_l U_{\mathrm{QSVT}}^l(M) U_{0} \right] \ket{0}
    \end{split}
\end{equation}
for $k\geq3$, where $U_{\mathrm{QSVT}}^e(M)$ and $U_{\mathrm{QSVT}}^o(M)$ yield the even and odd terms of Eq. \eqref{eq:JacobiAsPolys}, respectively, while $U_{\mathrm{QSVT}}^l(M)$ corresponds to a block encoding of $(-M)^k$ realized via QSVT. The exact form of the respective coefficients $c_e$, $c_o$ and $c_l$ is derived in the subsequent section and given in Eq. \eqref{eq:LCUJacobiCoefficients}.

Let us reiterate that the QSVT-enhanced quantum Jacobi algorithm requires the matrix $M$ to be Hermitian, a condition that imposes strong structural restrictions on $A$ and thus limits the general applicability of the approach. For real linear systems with $A\in \mathbb{R}^{N\times N}$, the algorithm can only be applied to matrices that admit a positive diagonal similarity that yields a symmetric matrix. This is equivalent to the element-wise requirement that for all
$i\neq j$ we have
\begin{equation} \label{eq:requirement}
    A_{ij}\cdot A_{jj} = A_{ji}\cdot A_{ii}.
\end{equation}

\subsection{Circuit implementation}

\begin{figure*}
    \centering
    \includegraphics[width=0.8\linewidth]{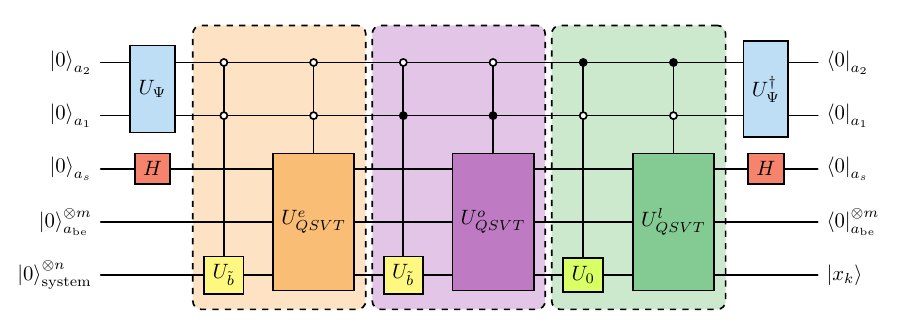}
    \caption{Circuit representation of the QSVT-enhanced Jacobi algorithm, where three QSVT subroutines are combined via LCU. The ancilla qubits $a_1$ and $a_2$ encode the LCU coefficients, $a_s$ denotes the signal qubit on which the phase rotations of the QSVT sequence are performed, and $a_{\text{be}}$ represents the ancilla register used for the block encoding of $M$.}
    \label{fig:QSVTCircuit}
\end{figure*}

The three QSVT subroutines from Eq. \eqref{eq:FinalJacobi} can be combined via the linear combination of unitaries (LCU), requiring two ancilla qubits in which the coefficients $c_e, c_o, c_l$ are encoded. 
The first QSVT subroutine $U_{\mathrm{QSVT}}^e(M)$ is intended to block encode the even exponents of $M$ from the Jacobi scheme, e.g. $I + M^2 + M^4 + ... + M^{k_e}$, where $k_e = 2(\lceil\frac{k}{2}\rceil-1)$ is the highest power of the even polynomial. However, for QSVT the polynomial has to obey $P(a)<1$ for $a\in[-1,1]$, which can be ensured by rescaling the polynomial. This can be done in two ways. One can evaluate $P(a)$ on the given interval and then divide the polynomial by its maximum value on that interval, which requires additional computation. A simple alternative is to divide the polynomial by the number of even terms that need to be considered for a given number of iterations, given by $m_e=\lceil\frac{k}{2}\rceil$. Furthermore, the block encoding requires the matrix $M$ to have $\norm{M}\leq1$, which can be guaranteed by rescaling $M$ if its norm is $>1$. Thus, we introduce the rescaling parameter $\alpha\coloneqq \max(1, \norm{M})$ and work with the normalized matrix $M/\alpha$. 
The same considerations can be done for the odd and the last part, corresponding to $U_{\mathrm{QSVT}}^o(M)$ and $U_{\mathrm{QSVT}}^l(M)$, respectively. For the odd part, the highest power is given by $k_o=2\lfloor\frac{k}{2}\rfloor-1$, and the Jacobi scheme includes $m_o = \lfloor \frac{k}{2} \rfloor$ odd terms. 

In total, the three polynomials for the QSVT subroutines are
\begin{equation}\label{eq:JacobiPolys}
    \begin{aligned}
    P_e(a) &= \frac{1}{m_e} \frac{1}{\alpha^{2(m_e-1)}} \sum_{j=0}^{m_e-1}
    \left(\alpha a\right)^{2j}, \\
    P_o(a) &= -\frac{1}{m_o} \frac{1}{\alpha^{2m_o-1}} \sum_{j=0}^{m_o-1}
    \left(\alpha a\right)^{2j+1}, \\
    P_l(a) &= (-1)^k a^k .
    \end{aligned}
\end{equation}
These definitions of the polynomials can be used to determine the corresponding QSVT angles for the single-qubit phase gates.
The rescaling factors of the polynomials, as well as those of the quantum states $\ket*{\Tilde{b}}$ and $\ket{x_0}$, can be incorporated into the LCU coefficients
\begin{equation}\label{eq:LCUJacobiCoefficients}
    \begin{split}
        c_e &= m_e \,\hat{b}\, \alpha^{2(m_e-1)}, \\
        c_o &= m_o\,\hat{b}\, \alpha^{2m_o-1},\\
        c_l &= \hat{x}_0 \, \alpha^{k}.
    \end{split}
\end{equation}
Thus, the LCU ancilla state can be chosen as
\begin{equation}
    \begin{split}
        \ket{\Psi}
            =&
            \frac{1}{\sqrt{c_e + c_o + c_l}}
            \left(
            \sqrt{c_e}\ket{00}
            +
            \sqrt{c_o}\ket{01} \right. \\
            & \left. +
            \sqrt{c_l}\ket{10}
            \right),
    \end{split}
\end{equation}
where the remaining basis state $\ket{11}$ is assigned zero amplitude. This state can be prepared by the unitary operator $U_\Psi$ acting on the ancilla register such that $U_\Psi \ket{00} = \ket{\Psi}$.

In our quantum Jacobi algorithm, the QSVT sequence is used as a subroutine controlled by the LCU ancilla qubits, as illustrated in Fig.~\ref{fig:QSVTCircuit}. It is therefore worth noting that in a controlled version of QSVT only the block-encoding unitary $U_M$ and the single-qubit $Z$-rotations need to be conditioned on the LCU ancillas. In contrast, the CNOT gates arising from the projector-controlled phase-shift operators do not require additional control by the LCU ancilla. 

Additionally, we emphasize that the polynomials in Eq. \eqref{eq:JacobiPolys} are real-valued. Since QSVT generally yields a complex polynomial transformation, we must ensure that only the real component is realized in order to correctly implement the Jacobi algorithm. We can implement a block encoding of a real polynomial by combining the polynomial with its complex conjugate, according to $P_\text{real}(a)=\frac{1}{2}(P(a)+P^\ast(a))$. Instead of implementing a separate circuit for the complex-conjugated part, the desired real polynomial can be efficiently realized by applying a Hadamard gate to the signal qubit $a_s$ \cite{lin_lecture_2022}. This technique allows an efficient implementation with respect to the circuit depth. 
After uncomputing the LCU ancilla register and the signal qubit, and conditioning on measuring all ancilla qubits in the state $\ket{0}$, the resulting state corresponds to the Jacobi approximation $\ket{x_k}$ of the solution.

\subsection{Resource estimation}

\begin{figure*}[t]
\centering
\begin{subfigure}{0.49\textwidth}
  \centering
  \includegraphics[width=\linewidth]{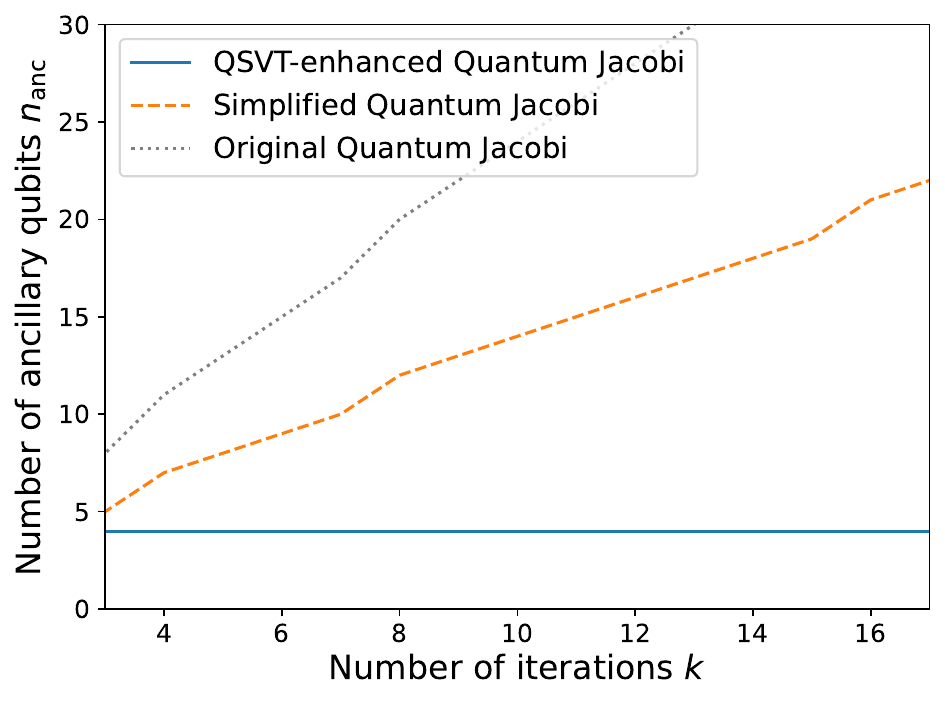}
  \caption{}
  \label{fig:WidthScaling}
\end{subfigure}
\hfill
\begin{subfigure}{0.49\textwidth}
  \centering
  \includegraphics[width=\linewidth]{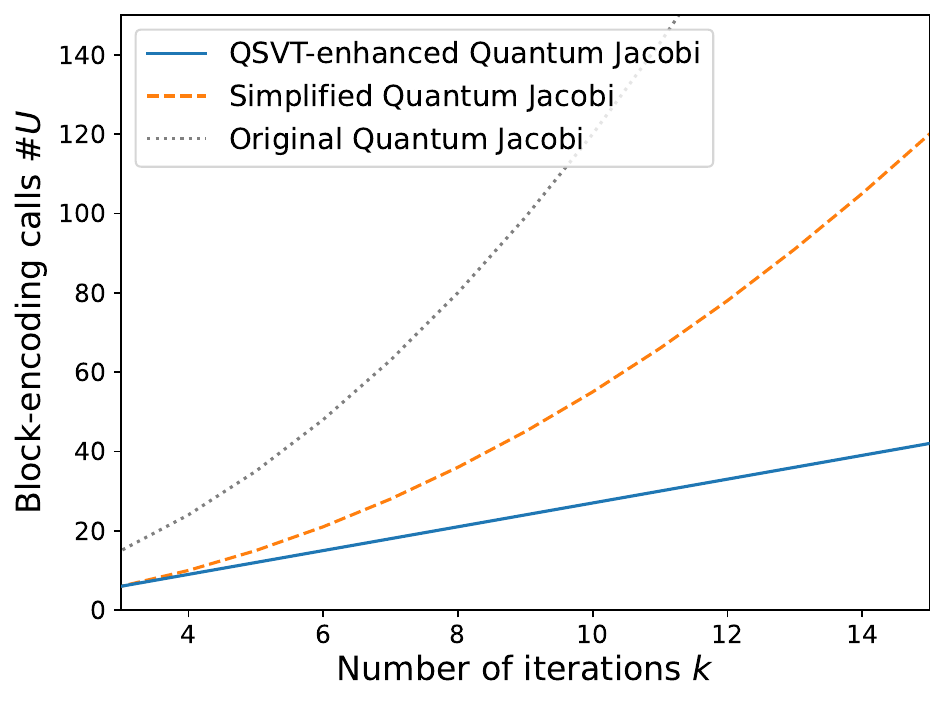}
  \caption{}
  \label{fig:DepthScaling}
\end{subfigure}
\caption{Comparison of qubit requirements and circuit depth for different quantum implementations of the Jacobi method. The number of ancillary qubits is shown in (a), comparing the QSVT-enhanced implementation [Eq. \eqref{eq:FinalJacobi}] with the simplified implementation based on naive block-encoding multiplications [Eq. \eqref{eq:ModifiedJacobi}] and the original implementation [Eq. \eqref{eq:OriginalJacobi}] from Ref. \cite{williams_quantum_2026}. The curves correspond to the illustrative minimal case with a single ancilla qubit per block encoding ($m=1$).
The circuit depth, given by the number of calls to the block encoding, is shown in (b).
}
\label{fig:Scaling}
\end{figure*}

Based on the circuit construction described above, we now analyse the required quantum resources for our proposed quantum Jacobi algorithm.

The total number of qubits consists of the system register, the two ancilla qubits for the LCU construction, a signal qubit, as well as additional ancilla qubits used for the block encoding. In this work, we do not consider a specific implementation of block encodings. Instead, we treat $U_M$ as a generic block-encoding unitary of the matrix $M$, which is realized using $m$ ancilla qubits, and keep track of resources in terms of the number of $U_M$ queries. Then, the Jacobi circuit can be realized using $\log_2(N)+ m + 3$ qubits.

Let us investigate the circuit depth in terms of calls to the block encoding $U_M$. Since $k_e$ and $k_o$ denote the highest powers of the even and odd polynomials, respectively, the corresponding QSVT subroutines require $k_e$ and $k_o$ calls to $U_M$, while the last QSVT subroutine requires $k$ calls. In total, the QSVT-based implementation of the quantum Jacobi algorithm requires $3k-3$ calls to the block encoding. Hence, the circuit depth scales linearly with the number of Jacobi iterations $k$. 

The original implementation, as described in Ref. \cite{williams_quantum_2026} and summarized in Sec. \ref{sec:JacobiIntro}, needs $k(k+3)/2$ calls to the block encoding $U_{D^{-1}}$ and $k(k+1)/2$ calls to the block encoding $U_{R}$, which sums up to $k(k+2)$ calls. However, to ensure a fair comparison, we also consider a modified version of Eq.~\eqref{eq:OriginalJacobi}. In this version, as in the QSVT-enhanced approach, we pre-compute $M = D^{-1} R$ and $\tilde{b} = D^{-1} b$ classically. Then
\begin{equation}\label{eq:ModifiedJacobi}
    \ket{x_k} = \left[ \sum_{j=1}^k \left( -U_M \right)^{j-1}U_{\Tilde{b}} + \left( -U_{M} \right)^{k} U_0 \right] \ket{0}
\end{equation}
can be implemented, in analogy to the original algorithm, using a combination of multiplications of block encodings and LCU. In this case, the corresponding circuit requires $\log_2(N)+k\cdot m+\lceil \log_2(k+1) \rceil$ qubits and $k(k+1)/2$ calls to the block encoding $U_M$. Hence, the naive implementation based on repeated block-encoding multiplications requires $\mathcal{O}(k^2)$ calls to the block encoding, while the QSVT-based implementation reduces this to $\mathcal{O}(k)$. As illustrated in Fig.~\ref{fig:DepthScaling}, the QSVT-enhanced Jacobi algorithm shows a clear improvement in both circuit depth and width. 

Finally, let us briefly discuss the success probability of the proposed algorithm. As for any LCU construction, it depends on the normalization factor given by the sum of the LCU coefficients \cite{childs_hamiltonian_2012}. Thus, the success probability is
\begin{equation}
    p_\text{succ}=\frac{\norm{\ket{x_k}}^2}{C(k, \alpha)^2}
\end{equation}
with $C(k, \alpha)=c_e+ c_0+c_l$. Since the coefficients in Eq. \eqref{eq:LCUJacobiCoefficients} scale with powers of the rescaling parameter $\alpha$, we have $C(k,\alpha)=\mathcal{O}(\alpha^k)$. Consequently, a rescaling with $\alpha>1$ would lead to an exponential suppression of the success probability.
However, the algorithm is restricted to linear systems whose iteration matrix $M$ is Hermitian. In this case, the operator norm is equal to the spectral radius \cite{meyer_matrix_2023}. The convergence condition of the Jacobi iteration therefore implies $\norm{M}=\rho(M)<1$ and consequently $\alpha=1$. As a result, the normalization factor simplifies to $C(k,1)=k \hat{b}+ \hat{x}_0$ and the success probability therefore scales as $p_\text{succ} \propto 1/k^2$. Hence, the probability of obtaining the desired output state decreases quadratically with the number of Jacobi iterations. As in other LCU-based quantum algorithms, this success probability can be further increased using amplitude amplification \cite{brassard_quantum_2002}.

\section{Application to Poisson Problems}\label{sec:Application}

In this section, we demonstrate the proposed quantum Jacobi algorithm for solving the Poisson equation. We first consider simple one-dimensional examples to illustrate its application to discretized Poisson problems. We then apply the algorithm to the pressure Poisson equation arising in lid-driven cavity flow, demonstrating its applicability to a physically relevant system.

\subsection{One-Dimensional Poisson Equation}

\begin{figure*}
\centering
\begin{subfigure}{.49\textwidth}
  \centering
  \includegraphics[width=\linewidth]{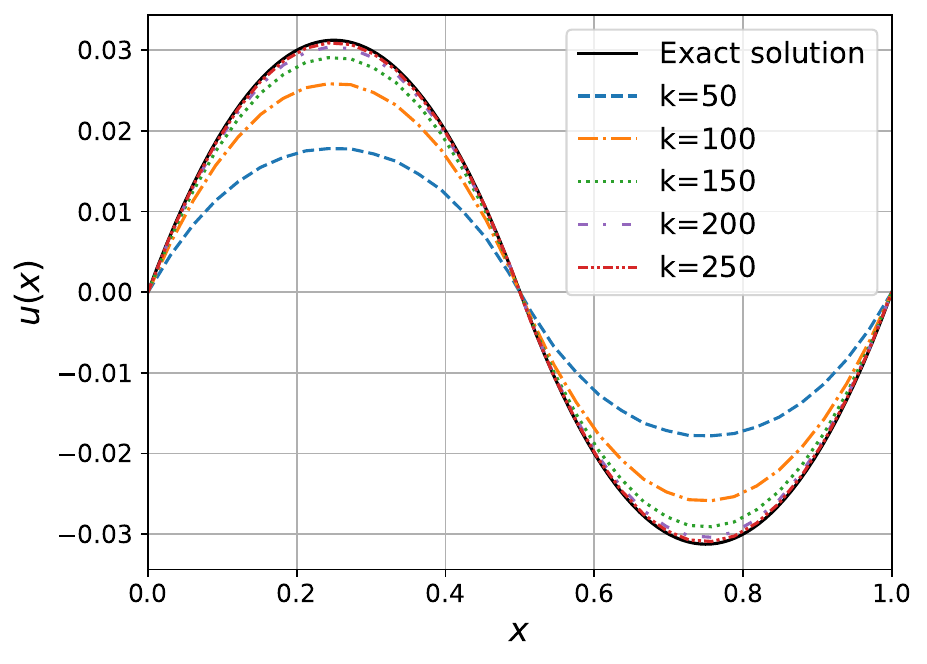}
  \caption{$f(x)=2\textup{H}(x-0.5)-1$, $u(0)=u(1)=0$}
  \label{fig:PoissonHeaviside}
\end{subfigure}%
\begin{subfigure}{.49\textwidth}
  \centering
  \includegraphics[width=\linewidth]{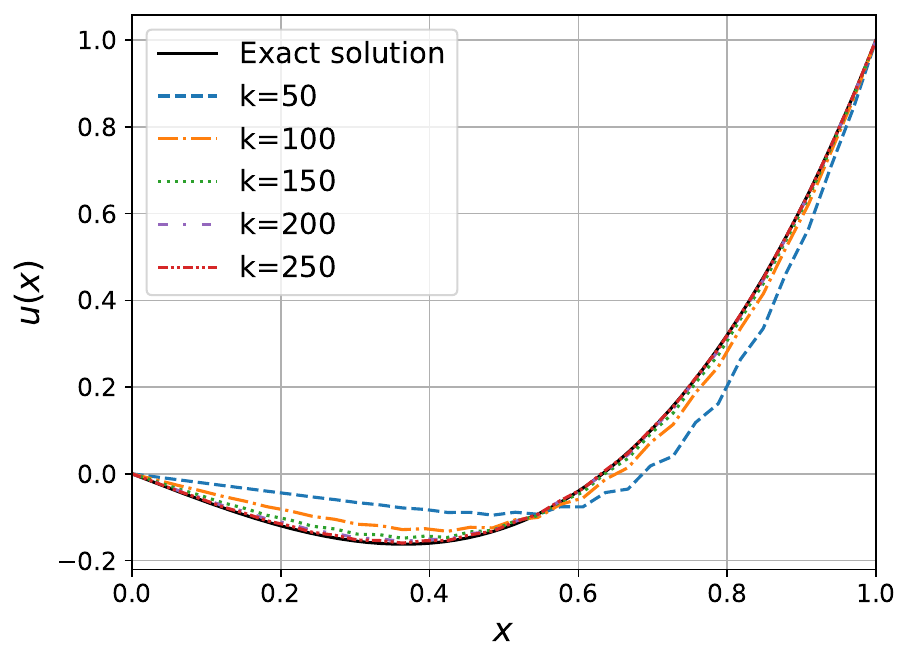}
  \caption{$f(x)=10x$, $u(0)=0$, $u(1)=1$}
  \label{fig:PoissonLinear}
\end{subfigure}
\caption{Solutions of the one-dimensional Poisson equation $\frac{d^2u}{dx^2}=f(x)$ obtained with the quantum Jacobi algorithm for different source functions. Results are shown for different numbers of Jacobi iterations $k$ on a uniform grid with $N=32$ interior points.}
\label{fig:Poisson1Dexamples}
\end{figure*}

As a first demonstration of the quantum Jacobi algorithm, we consider the one-dimensional Poisson equation on the unit interval
\begin{equation}
    \frac{d^2u}{dx^2}=f(x), \quad x \in [0,1]
\end{equation}
with Dirichlet boundary conditions $u(0)=\beta$, $u(1)=\gamma$, and $\beta,\gamma\in\mathbb{R}$.
A second-order central finite-difference discretization on a uniform grid with spacing $\Delta x$ yields the linear system $Au=b$ with
\begin{equation}
    A =
    \begin{pmatrix}
    -2 & 1 &        &        &   \\
    1  & -2 & 1     &        &   \\
       & \ddots & \ddots & \ddots &   \\
       &        & 1 & -2 & 1 \\
       &        &   & 1 & -2
    \end{pmatrix},
    \,
    b = \Delta x^2
    \begin{pmatrix}
    f_1-\beta\\
    f_2\\
    \vdots\\
    f_{N-1}\\
    f_N-\gamma
    \end{pmatrix},
    \label{eq:1DPoissonSystem}
\end{equation}
where the Dirichlet boundary conditions are incorporated into the right-hand side. The linear system satisfies the requirement of the QSVT-enhanced quantum Jacobi algorithm given in Eq. \eqref{eq:requirement}. 

We consider two representative benchmark problems following Ref. \cite{ghafourpour_applicability_2025}. The first employs the discontinuous source term $f(x)=2\textup{H}(x-0.5)-1$, where $\textup{H}$ denotes the Heaviside function, together with homogeneous boundary conditions $u(0)=u(1)=0$. The second considers the linear source term $f(x)=10x$, with inhomogeneous boundary conditions $u(0)=0$ and $u(1)=1$. 

The solutions obtained using the quantum Jacobi algorithm are shown in Figure \ref{fig:Poisson1Dexamples} for different numbers of Jacobi iterations $k$. 
All numerical simulations of the quantum circuits were performed using the \textsf{PennyLane} package \cite{bergholm_pennylane_2022} and it's function \lstinline{qml.poly_to_angles} was used to compute the QSVT phase angles. As expected, the approximation improves with increasing iteration number and converges towards the analytical solution. Furthermore, the quantum implementation exactly reproduces the classical Jacobi solution up to machine precision, with maximum deviations below $<10^{-13}$. 

In general, the convergence of the Jacobi method depends on the choice of the initial guess. For the present examples, we use $x_0=b$. In practical applications, however, a more accurate initial guess is often available. For example, in the lid-driven cavity simulation considered below, the solution from the previous time step naturally provides an improved starting point for the iteration.

\subsection{Solving the Poisson equation in Lid-driven cavity flow}

For incompressible flow simulations, the Navier–Stokes equations are commonly solved using projection methods or pressure-correction methods such as SIMPLE \cite{ferziger_computational_2020}. These approaches enforce incompressibility by solving a pressure Poisson equation, which ensures that the velocity field is divergence-free. The resulting linear systems are large and sparse, and solving the pressure Poisson equation is computationally expensive, representing the dominant computational cost of these methods \cite{guermond_overview_2006, aithal_fast_2020, fang_fast_2023}. 

\begin{figure*}
\centering
\begin{subfigure}{.5\textwidth}
  \centering
  \includegraphics[width=.9\linewidth]{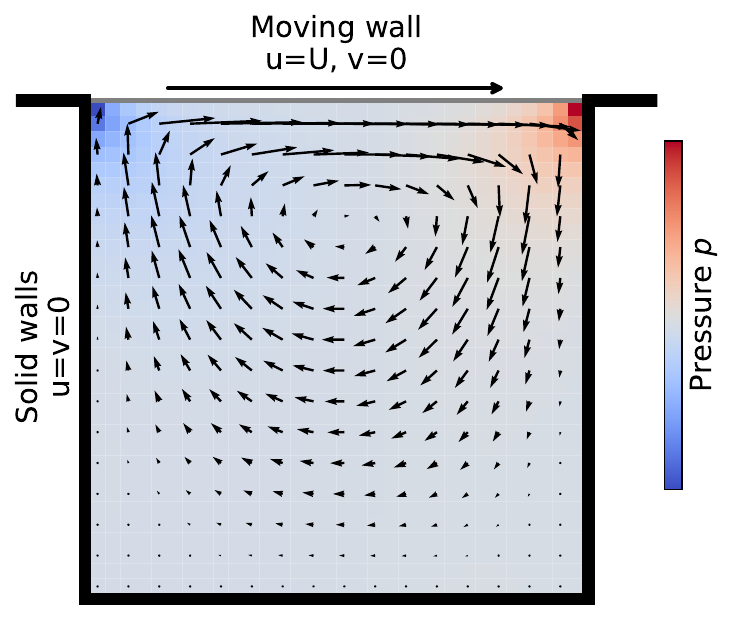}
  \caption{}
  \label{fig:CavitySchematic}
\end{subfigure}%
\begin{subfigure}{.5\textwidth}
  \centering
  \includegraphics[width=.9\linewidth]{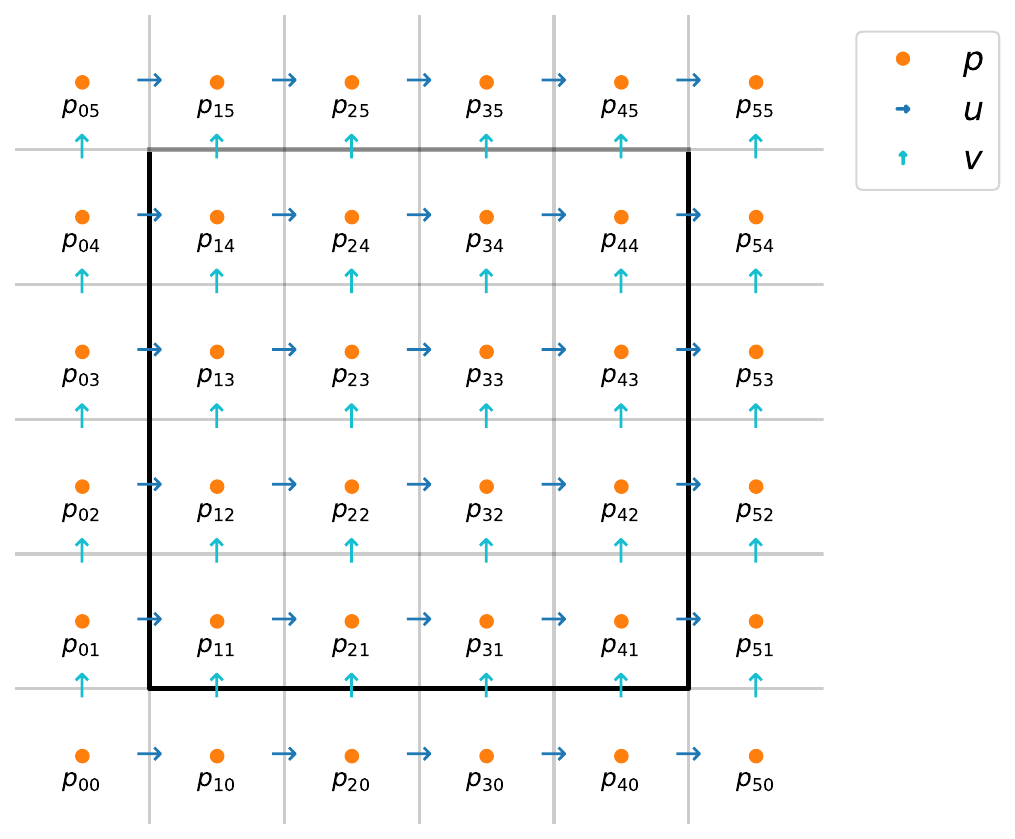}
  \caption{}
  \label{fig:CavityGrid}
\end{subfigure}
\caption{A schematic of the lid-driven cavity flow is shown in (a). The corresponding staggered grid used for the finite-difference discretization is shown in (b), where pressure values are located at the cell centers and the velocity components at the cell faces.}
\label{fig:Cavity}
\end{figure*}

In this work, we therefore consider the pressure Poisson equation arising in Chorin's projection method \cite{chorin_numerical_1968} as a physically motivated test problem to study our quantum Jacobi algorithm. In particular, we focus on the lid-driven cavity flow, a widely used benchmark in computational fluid dynamics \cite{ghia_high-re_1982} that has also been studied in the context of quantum algorithms, for example in Refs. \cite{lapworth_hybrid_2022, song_incompressible_2025, lee_multiple-circuit_2026}. It describes the flow inside a square cavity, that is induced by a moving top lid, as illustrated in Fig.~\ref{fig:CavitySchematic}. 

In Chorin's projection method, the Navier Stokes equation is discretized in time by the explicit Euler scheme 
\begin{equation}
    \begin{split}
        \frac{\vb{u}^{(n_t+1)}-\vb{u}^{(n_t)}}{\Delta t} =& -\frac{1}{\rho} \nabla p^{(n_t+1)} - \left( \vb{u}^{(n_t)} \cdot \nabla \right) \vb{u}^{(n_t)} \\
        &+ \nu \nabla^2 \vb{u}^{(n_t)},
    \end{split}
\end{equation}
where $\vb{u}^{(n_t)} = (u^{(n_t)}, v^{(n_t)})$ is the velocity field at the $n_t$-th time step, $\Delta t$ is the time step size, $\rho$ is the fluid density, $p$ is the pressure field and $\nu$ is the viscosity.
Then the equation is split up and solved in three steps:
\begin{enumerate}
    \item An intermediate velocity $\vb{u}^\ast$ is introduced by solving the equation without the pressure gradient term
    \begin{equation}
        \frac{\vb{u}^\ast-\vb{u}^{(n_t)}}{\Delta t} = - \left( \vb{u}^{(n_t)} \cdot \nabla \right) \vb{u}^{(n_t)} + \nu \nabla^2 \vb{u}^{(n_t)}.
    \end{equation}
    \item Next, the incompressibility constraints are considered with the pressure Poisson equation
    \begin{equation}\label{eq:PressurePoisson}
        \nabla^2 p^{(n_t+1)} = \frac{\rho}{\Delta t} \nabla \vb{u}^\ast.
    \end{equation}
    \item Finally, the velocity is updated by correcting the intermediate velocity
    \begin{equation}
        \vb{u}^{(n_t+1)} = \vb{u}^\ast -\frac{\Delta t}{\rho} \nabla p^{(n_t+1)}.
    \end{equation}
\end{enumerate}
In the second step, the pressure Poisson equation gives rise to a linear system of equations. We now turn to its discretization for the considered test case.

\subsubsection{Pressure Poisson equation}

The computational domain is discretized using a uniform staggered grid (see Fig.~\ref{fig:CavityGrid}), where the pressure values are stored at the cell centers, while the velocity components are stored at the cell surfaces. 
Furthermore we introduce ghost cells around the computational domain, to treat the boundary conditions. The inner computational domain is an $n_x \times n_y$ grid. Here, we consider a square grid with $n_x=n_y$ and uniform grid spacing $\Delta x=\Delta y$.
We apply Neumann boundary conditions at the left, right and bottom wall, while the moving top wall has Dirichlet boundary conditions. Accordingly, the ghost cell values satisfy $p_{0,j}=p_{1,j}$, $p_{n_x+1,j}=p_{n_x,j}$, $p_{i,0}=p_{i,1}$, and $p_{i,n_x+1}=-p_{i,n_x}$ for $i,j=1,\ldots,n_x$.

In two dimensions the pressure Poisson equation from Eq. \eqref{eq:PressurePoisson} reads
\begin{equation}\label{eq:PressurePoisson2D}
    \nabla^2 p = \frac{\rho}{\Delta t} \left( \frac{\partial u}{\partial x} + \frac{\partial v}{\partial y} \right).
\end{equation}
The Laplacian can be approximated using the five-point stencil finite-difference method, such that
\begin{equation}\label{eq:FivePointStencil}
    \begin{split}
        \nabla^2 p =& \frac{1}{\Delta x^2} \left( p_{i+1,j} + p_{i-1,j} \right. \\
        &+ \left( p_{i,j+1} + p_{i,j-1} - 4p_{i,j} \right).
    \end{split}
\end{equation}
We define the right hand side of Eq. \eqref{eq:PressurePoisson2D} as $d_{i,j}$.
Now, we transform the pressure Poisson equation into a linear system $Ax=b$. Vectorizing the pressure field $p$ and the right hand side $d$ yields the vectors $x,b\in\mathbb{R}^N$, where $N=n_x\cdot n_y$.
The matrix $A\in\mathbb{R}^{N\times N}$ has the block form
\begin{equation}\label{eq:PressureMatrix}
    A = 
    \begin{pmatrix}
        A_b & I & & & \\
        I & A_c & I & & \\
         & \ddots & \ddots & \ddots & \\
         & & I & A_c & I \\
         & & & I & A_t
    \end{pmatrix}
\end{equation}
with
\begin{equation}
    \begin{split}
        A_b &= 
        \begin{pmatrix}
            -2 & 1 & & & \\
            1 & -3 & 1 & & \\
             & \ddots & \ddots & \ddots & \\
             & & 1 & -3 & 1 \\
             & & & 1 & -2
        \end{pmatrix}, \\
        A_c &= 
        \begin{pmatrix}
            -3 & 1 & & & \\
            1 & -4 & 1 & & \\
             & \ddots & \ddots & \ddots & \\
             & & 1 & -4 & 1 \\
             & & & 1 & -3
        \end{pmatrix}, \\
        A_t &= 
        \begin{pmatrix}
            -4 & 1 & & & \\
            1 & -5 & 1 & & \\
             & \ddots & \ddots & \ddots & \\
             & & 1 & -5 & 1 \\
             & & & 1 & -4
        \end{pmatrix},
    \end{split}
\end{equation}
where $A_b,A_c,A_t\in\mathbb{R}^{n_x\times x_x}$, $I\in\mathbb{R}^{n_x\times x_x}$ is the identity matrix, and the boundary conditions are incorporated into the matrix.

\subsubsection{Treatment of boundary conditions}\label{sec:TreatmentBoundary}
Considering our testcase of the pressure Poisson equation, the matrix given in Eq. \eqref{eq:PressureMatrix} for the linear system does not satisfy the requirement of Eq. \eqref{eq:requirement} for our quantum Jacobi algorithm.
From considering Eq. \eqref{eq:FivePointStencil} it becomes clear that the five point stencil corresponds to the symmetric Laplace matrix
\begin{equation}\label{eq:LaplaceMatrix}
    L = 
    \begin{pmatrix}
        L_c & I & & & \\
        I & L_c & I & & \\
         & \ddots & \ddots & \ddots & \\
         & & I & L_c & I \\
         & & & I & L_c
    \end{pmatrix}
\end{equation}
with
\begin{equation}
    L_c= 
    \begin{pmatrix}
        -4 & 1 & & & \\
        1 & -4 & 1 & & \\
         & \ddots & \ddots & \ddots & \\
         & & 1 & -4 & 1 \\
         & & & 1 & -4
    \end{pmatrix}
\end{equation}
and that the non-symmetric entries in $A$ arise due to the boundary conditions.
So instead of solving the linear system $Ax=b$, we will instead solve
\begin{equation}\label{eq:ModifiedLinearSys}
    Lx^{(n_t)} = b-Cx^{(n_t-1)},
\end{equation}
where $C=A-L$ contains the deviations imposed by the boundary conditions. This induces an error that is proportional to $\norm{x^{(n_t)}-x^{(n_t-1)}}$, as we now take the solution from the previous pseudo time step for this part. This error decreases during Chorin's projection method, as the solution approaches convergence and the difference between consecutive pressure fields becomes smaller. The practical impact of this approximation is investigated in the following subsection.  \\

\subsubsection{Numerical results}

\begin{figure*}
\centering
\begin{subfigure}{.57\textwidth}
  \centering
  \includegraphics[width=\linewidth]{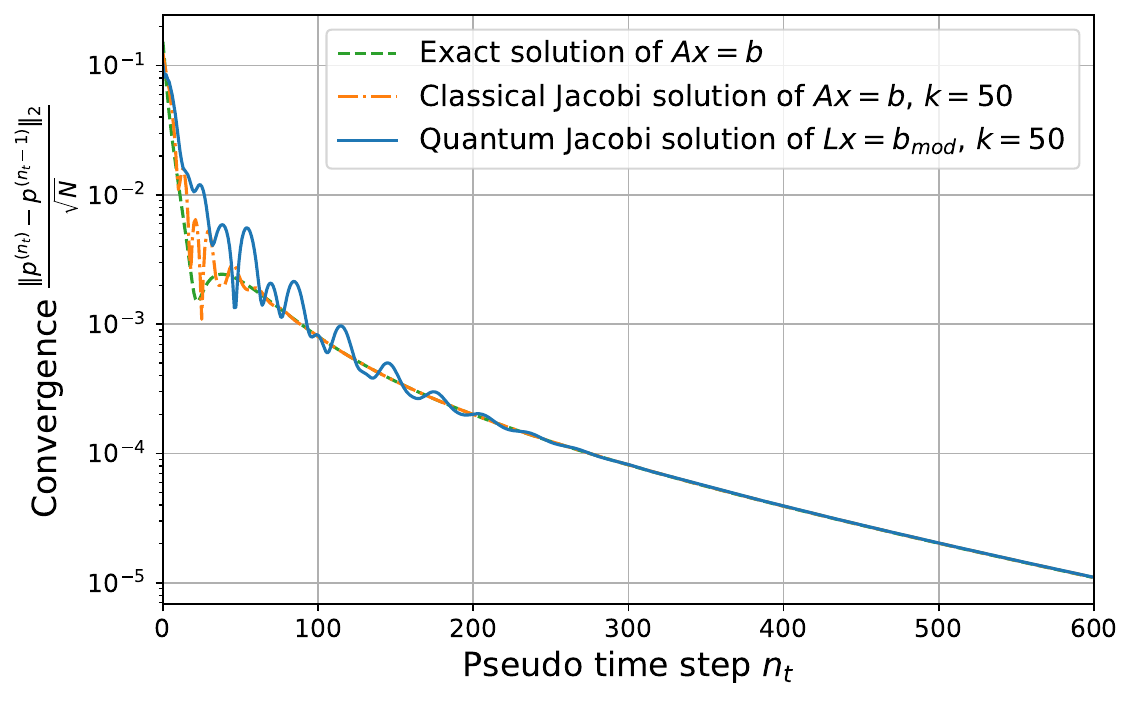}
  \caption{}
  \label{fig:LidConvergence}
\end{subfigure}%
\begin{subfigure}{.43\textwidth}
  \centering
  \includegraphics[width=\linewidth]{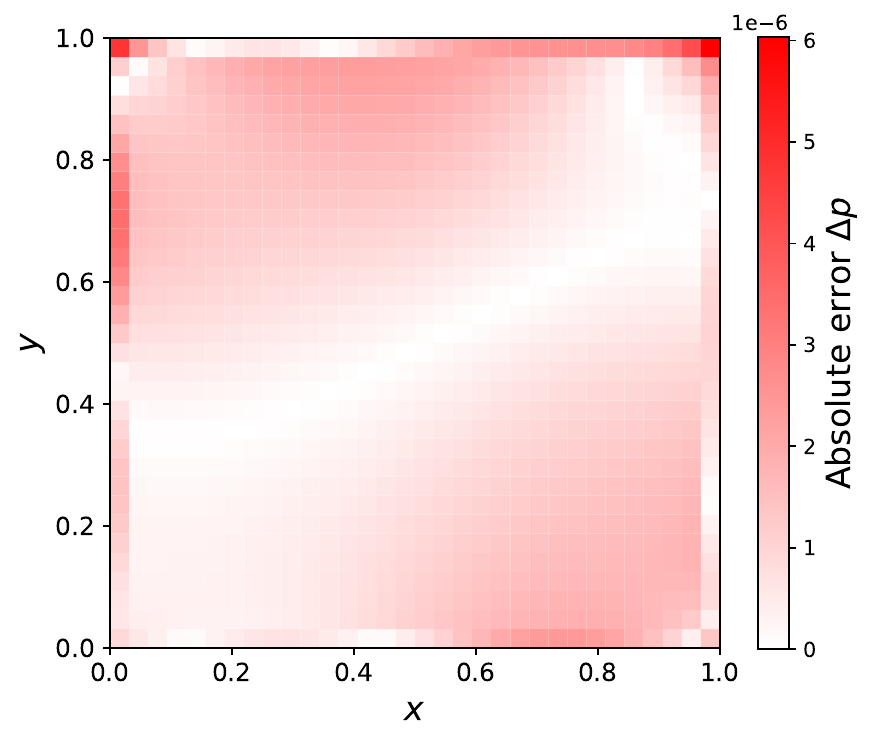}
  \caption{}
  \label{fig:LidError}
\end{subfigure}
\caption{Comparison of solving the symmetrized linear system from Eq. \eqref{eq:ModifiedLinearSys} using the proposed quantum Jacobi algorithm with the classical solution of the original linear system. The convergence of Chorin's projection method is shown in (a). Subfigure (b) shows the absolute difference between the pressure field obtained with the quantum Jacobi algorithm and the pressure field obtained by solving the original linear system exactly at pseudo time step $n_t=600$.}
\label{fig:LidResults}
\end{figure*}

Solving the pressure Poisson equation to high accuracy requires a large number of Jacobi iterations, which increases with the grid resolution. Here, we consider grids of size $n_x \times n_x$ with $n_x \in {4,8,16,32}$. For the largest grid ($n_x=32$), several thousand Jacobi iterations would be required to fully converge the linear system. In practice, however, Chorin's projection method converges even when the pressure Poisson equation is solved only approximately, making already a low number of iterations, such as $k=50$, sufficient. 

Quantum circuit simulations were performed for up to $k=80$ Jacobi iterations. For all considered grid sizes and iteration numbers, the quantum Jacobi algorithm again reproduces the classical Jacobi iterates up to machine precision, with maximum deviations below $10^{-12}$. Consequently, the polynomial approximation and the computed QSVT phase angles do not introduce any noticeable numerical error. 

To evaluate the practical impact of the symmetric approximation introduced in Section \ref{sec:TreatmentBoundary}, we compare the resulting CFD simulations with those obtained from the original pressure Poisson equation. Figure \ref{fig:LidConvergence} compares the convergence of Chorin's projection method using the proposed quantum Jacobi algorithm with the reference solution obtained by solving the original linear system exactly and with the classical Jacobi method using $k=50$ iterations.
During the first pseudo time steps, differences in the convergence behaviour are observed due to the approximation of the linear system and the modified treatment of the boundary conditions. As the projection method converges, however, the convergence histories become identical, confirming that $\norm{x^{(n_t)}-x^{(n_t-1)}}$ decreases during the simulation. 

The resulting influence on the pressure solution is illustrated in Figure~\ref{fig:LidError}, which shows the absolute difference between the pressure field obtained using the quantum Jacobi algorithm for the modified linear system and the reference solution obtained by solving the original system after $600$ pseudo time steps. The error is of the order of $10^{-6}$ and is mainly localized near the boundaries. This demonstrates that the approximation introduced to satisfy the requirements of the QSVT-enhanced quantum Jacobi algorithm has only a minor influence on the resulting pressure field.

\section{Discussion of Polynomial-Based Quantum Implementations of Iterative Methods}\label{sec:Discussion}

As we have seen, the use of QSVT subroutines for the quantum Jacobi algorithm enables an efficient implementation of the iteration scheme, but relies on the normality of the underlying operator. This requirement already restricts the class of linear systems for which our quantum Jacobi implementation can be applied. 

Beyond the Jacobi method, several classical stationary iterative schemes exhibit improved convergence properties, most notably the Gauss-Seidel and successive over-relaxation (SOR) methods. It is therefore natural to investigate whether such methods can be translated to the quantum setting in a similar manner. 

We have studied the quantum implementation of the Gauss-Seidel method as a representative example, for which the details are provided in the appendix \ref{sec:GaussSeidel}. However, in contrast to the Jacobi case, the operators arising from the Gauss-Seidel iteration are non-normal by construction. As a consequence, the corresponding matrix polynomials cannot be implemented with a QSVT framework. 

Both the restricted applicability of the quantum Jacobi method and the inability to extend the approach to more advanced iterative schemes arise from the normality requirement of QSVT in order to transform eigenvalues.
We therefore briefly review recent quantum frameworks that enable eigenvalue transformations of non-normal matrices.

One of the first general frameworks for eigenvalue transformation of non-normal matrices is the recently proposed Quantum Eigen-Value Transformation (QEVT) \cite{low_quantum_2024}.
The algorithm represents the target polynomial in a Chebyshev basis and constructs the corresponding Chebyshev history state. By performing a fixed-point amplitude amplification on part of the history state, it prepares a quantum state proportional to $p\left( \frac{M}{\alpha} \right) \ket{\psi}$. The preparation of the history state is formulated as a Quantum Linear System Problem. Solving the QLSP with an optimal QLSS yields the required Chebyshev history state.
The requirement of an optimal QLSS for the preparation of the Chebyshev history state currently limits the practical applicability of the QEVT algorithm. For this reason, we do not pursue a QEVT-based implementation in the present work. 

Yet another promising approach for realizing matrix polynomials is based on the Fast One-Qubit Controlled Select LCU (FOQCS-LCU) framework. This method enables the implementation of a block-encoding of a degree $d$ polynomial with a circuit depth that scales only linearly in $d$, while the additional cost of the polynomial transformation remains independent of the complexity of the underlying block-encoding of the matrix \cite{nibbi_practical_2026}. The approach relies on a decomposition of the operator into Pauli strings, exploiting their structured representation in terms of $X$- and $Z$-components. Consequently, it is particularly efficient for matrices that admit a sparse and structured Pauli decomposition. 

Most recently, Gutiérrez et al. \cite{gutierrez_quantum_2026} extended Generalized Quantum Signal Processing (QGSP) to eigenvalue transformation of arbitrary matrices by introducing the concept of an $r$-regular block encoding\footnote{Referred to as $n$-regular block encoding in Ref.~\cite{gutierrez_quantum_2026}. Here, we use $r$ instead to avoid confusion with our notation, where $n$ denotes the number of qubits.}. An $r$-regular block encoding is constructed such that its $d$-th power reproduces the $d$-th power of the encoded matrix for every $0 \le d \le r$. The construction is based on the use of a quantum incrementor and requires $\mathcal{O}(r)$ ancilla qubits. 

As this framework became available only after the development of the present algorithm, we restrict ourselves to a brief discussion of its implications. Replacing the QSVT-based polynomial implementation of the Jacobi algorithm by GQSP acting on an $r$-regular block encoding would remove the constraints on the linear system of the present approach at the cost of an additional $\log(k)$ ancilla qubits. In the present work, we focus on the QSVT-based implementation, which remains the more resource-efficient approach whenever the iteration matrix is Hermitian. 

Together, these recent developments provide promising directions for extending quantum implementations of iterative schemes beyond the Jacobi method.

\section{Conclusion}

In this work, we have proposed an efficient quantum implementation of the iterative Jacobi method for solving systems of linear equations arising from the discretization of differential equations. The algorithm is based on a polynomial representation of the Jacobi iteration and, for Hermitian iteration matrices, uses the efficient implementation of polynomial transformations of block-encoded operators provided by QSVT. As a result, we achieved a favorable logarithmic scaling in the number of qubits with the system size while requiring only a constant ancilla overhead. The circuit depth scales linearly with the number of Jacobi iterations, reflecting the iterative nature of the algorithm. Finally, we demonstrated its applicability for one- and two-dimensional Poisson problems, including the pressure Poisson equation arising in Chorin's projection method for the lid-driven cavity flow. In all cases, the simulated quantum algorithm reproduced the classical Jacobi solutions with numerical precision. 

The present work demonstrated the potential of polynomial representations for quantum implementations of stationary iterative methods. In contrast to inversion-based quantum linear system solvers, the proposed algorithm directly implemented the classical Jacobi scheme, resulting in an approach that is more closely aligned with the solution strategies employed in modern CFD. The approach may therefore be well suited for integration into larger CFD workflows, where it could serve as a building block within more advanced solution frameworks. Realizing such workflows, however, requires efficient techniques for preparing and extracting quantum information, which remains an important challenge. 

Furthermore, the explicit circuit construction enabled a detailed assessment of the required quantum resources beyond asymptotic complexity estimates, indicating that the proposed algorithm is well suited to early fault-tolerant quantum computers, provided that efficient block encodings are available. Constructing such block encodings for practically relevant CFD operators, however, remains one of the central open challenges. 

Future work should increasingly focus on quantum algorithms that integrate naturally into existing CFD workflows rather than treating the solution of a linear system as an isolated matrix inversion problem. In particular, extending quantum iterative methods to multigrid algorithms or preconditioning techniques represents a promising direction for future research.

\appendix

\section{Quantum Singular Value Transformation}\label{sec:QSVT_Details}

The following sections provide a detailed treatment of QSVT, including its origin from quantum signal processing (QSP), the associated polynomial constraints, and the connection between singular-value and eigenvalue transformations.

\subsection{Quantum Signal Processing}

QSP allows the implementation of polynomial transformations of a scalar signal $a\in [-1,1]$ as a matrix element $P(a)=\expval{U_{\Vec{\phi}}}{0}$ \cite{martyn_grand_2021}.
Given the signal operator
\begin{equation}
    U_a = 
    \begin{pmatrix}
        a & i\sqrt{1-a^2} \\
        i\sqrt{1-a^2} & a
    \end{pmatrix} = e^{i \arccos(a) \sigma_x}.
\end{equation}
and a sequence of phase angels $\Vec{\phi}=(\phi_0, \phi_1, ...,\phi_d) \in \mathbb{R}^{d+1}$, the QSP sequence $U_{\Vec{\phi}}$ produces the matrix
\begin{equation}
    \begin{split}
    U_{\Vec{\phi}} &= e^{i\phi_0\sigma_z} \prod_{j=1}^d U_a e^{i\phi_j \sigma_z} \\
    &= 
    \begin{pmatrix}
        P(a) & iQ(a)\sqrt{1-a^2} \\
        iQ^\ast(a)\sqrt{1-a^2} & P^\ast(a)
    \end{pmatrix},
    \end{split}
\end{equation}
where $P,Q \in\mathbb{C}[a]$ satisfy
\begin{enumerate}[(i)]
        \item $\deg(P) \leq d$, $\deg(Q) \leq d-1$,
        \item $P$ has parity $(d \mod 2)$, $Q$ has parity $(d-1 \mod 2)$, and
        \item $\forall a\in [-1,1]:\,|P(a)|^2 + (1-a^2)|Q(a)|^2 = 1.$
    \end{enumerate}
It is useful to change the so-called signal-basis to be $\ket{+},\ket{-}$ and define $\text{Poly}(a)=\expval{U_{\Vec{\phi}}}{+} = \text{Re}[P(a)]+ i\text{Re}[Q(a)]\sqrt{1-a^2}$. In this case any polynomial with parity $d \mod 2$, $\deg(\text{Poly}) \leq d$ and $|\text{Poly}(a)|\leq 1 \, \forall a \in [-1,1]$ can be approximated.

\subsection{Quantum Eigenvalue Transformation of Hermitian matrices}\label{sec:QSVT_Details_QET}

The methodology of QSP can be extended to Hermitian operators, such as Hamiltonians, through Quantum Eigenvalue Transformation (QET) \cite{martyn_grand_2021}. \\
Consider a Hermitian matrix $H$ with it's eigendecomposition $H=\sum_k \lambda_k \dyad{v_k}$ as well as the corresponding block encoding
\begin{equation}
    U_H =
    \begin{blockarray}{ccc}
     & \mbox{\scriptsize $\Pi$} & \\
     \begin{block}{c(cc)}
         \mbox{\scriptsize $\Pi\;\;$} & H & \cdot\, \\
         & \cdot & \cdot\, \\
     \end{block}
     \end{blockarray} \;\; ,
\end{equation}
where the location of $H$ is determined by the projector $\Pi$. Let 
\begin{equation}
    \Pi_\phi\coloneqq e^{i \phi (2\Pi - I)}
\end{equation}
be the projector-controlled phase-shift operator. Then, the sequence
\begin{equation}
    \begin{split}
        U_{\Vec{\phi}} &=
        \begin{cases}
            \Pi_{\phi_{0}} \left[ \prod_{j=1}^{d/2} U_H^\dag \Pi_{\phi_{2j-1}} U_H \Pi_{\phi_{2j}} \right], \\
            \hfill \text{for even}\,\, d \\[0.4em]
            \Pi_{\phi_{0}} U_H \Pi_{\phi_{1}} \left[ \prod_{j=1}^{(d-1)/2} U_H^\dag \Pi_{\phi_{2j}} U_H \Pi_{\phi_{2j+1}} \right], \\
            \hfill \text{for odd}\,\, d
        \end{cases} \\
        &= 
        \begin{blockarray}{ccc}
         & \mbox{\scriptsize $\Pi$} & \\
         \begin{block}{c(cc)}
            \mbox{\scriptsize $\Pi\;\;$} & \textup{Poly}(H) & \cdot\, \\
             & \cdot & \cdot\, \\
         \end{block}
        \end{blockarray}
    \end{split}
\end{equation}
is a polynomial transformation of the eigenvalues of $H$ with
\begin{equation}
    \textup{Poly}(H) = \sum_k \textup{Poly}(\lambda_k)\dyad{v_k},
\end{equation}
where $\textup{Poly}(H)$ satisfies the QSP conditions introduced above.

Since Hermitian matrices are normal, their eigendecomposition can be written as $H=V \Lambda V^{-1}$ with a unitary matrix $V$. The unitary eigendecomposition of Hermitian matrices enables polynomial transformation of their eigenvalues on quantum computers. 

\subsection{Quantum Singular Value Transformation}

Although a unitary eigendecomposition is not available for general matrices, every matrix admits a singular value decomposition (SVD)
\begin{equation}
    M= W \Sigma V^\dag =\sum_k \sigma_k \op{w_k}{v_k},
\end{equation}
where $\Sigma$ contains the singular values $\sigma_k\geq0$, and where the left and right singular vectors form the unitary matrices $W$ and $V$.

The QSVT construction presented in Section \ref{sec:QSVT} then implements a polynomial transformation of the singular values \cite{gilyen_quantum_2019, martyn_grand_2021}, according to 
\begin{equation}
    \textup{Poly}(M)=
    \begin{cases}
        \sum_k \textup{Poly}(\sigma_k) \dyad{v_k},  \quad\text{for even}\,\, d \\[0.4em]
        \sum_k \textup{Poly}(\sigma_k) \op{w_k}{v_k},  \quad\text{for odd}\,\, d
    \end{cases},
\end{equation}
where $\textup{Poly}(M)$ again satisfies the QSP conditions introduced above.

\subsection{Relation between QET and QSVT}

Although the quantum Jacobi algorithm considered in this work relies on polynomial transformations of eigenvalues, we formulate the discussion in terms of QSVT. Therefore, it is useful to briefly review the relation between QET and QSVT. When the Quantum Singular Value Transformation is applied to Hermitian matrices the circuit is the same, as the one for the Quantum Eigenvalue Transformation. 
This can be easily shown, as the singular values and eigenvalues satisfy $\sigma_k=|\lambda_k|$ for Hermitian matrices. Thus
\begin{equation}
    \begin{split}
        H&=\sum_k \lambda_k\op{v_k}{v_k} = \sum_k  |\lambda_k| \op{\text{sign}(\lambda_k)v_k}{v_k} \\
        &= \sum_k \sigma_k \op{w_k}{v_k}.
    \end{split}
\end{equation}
For an even polynomial
\begin{equation}
    \begin{split}
        \text{Poly}(H) &= \sum_k \text{Poly}(\lambda_k) \dyad{v_k} \\
        &= \sum_k \text{Poly}(|\lambda_k|) \dyad{v_k},
    \end{split}
\end{equation}
and for an odd polynomial
\begin{equation}
    \begin{split}
        \text{Poly}(H) &= \sum_k \text{Poly}(\lambda_k) \dyad{v_k} \\
        &= \sum_k \text{Poly}(|\lambda_k|) \op{w_k}{v_k}.
    \end{split}
\end{equation}
Furthermore, any non-Hermitian matrix $M$ can be extended to the dilated Hermitian matrix
\begin{equation}
    \Tilde{M} =
    \begin{pmatrix}
        0 & M^\dagger \\
        M & 0
    \end{pmatrix}.
\end{equation}
If this Hermitian matrix is now used for the Quantum Eigenvalue Transformation the result will be the same as for the QSVT. Consequently, polynomial eigenvalue transformations of Hermitian matrices can be viewed as a special instance of QSVT. For this reason, and to emphasize the limitations imposed by the hermiticity requirement discussed in the main text, we use the term QSVT throughout this work.

\section{Quantum Implementation of the Gauss-Seidel method}\label{sec:GaussSeidel}

In this section, we investigate whether the QSVT-based approach developed for the quantum Jacobi algorithm can be extended to the Gauss-Seidel method.

\subsection{Classical Gauss-Seidel method}

The classical Gauss–Seidel method is defined by the iterative update rule \cite{saad_iterative_2003}
\begin{equation}
x_k^{(i)} = \frac{1}{a_{ii}}\left(b_i - \sum_{j<i} a_{ij} x_k^{(j)} - \sum_{j>i} a_{ij} x_{k-1}^{(j)}\right),
\end{equation}
where $x_k^{(i)}$ is the $i$-th component of the solution after $k$ iterations.
In contrast to the Jacobi method, newly computed components $x_k^{(j)}$ with $j<i$ are used immediately within the same iteration. 

To express the method in matrix form, we decompose the system matrix as $A = D + B + T$, where $D$ denotes the diagonal, $B$ the strictly lower triangular, and $T$ the strictly upper triangular part of $A$, yielding
\begin{equation}
x_k = (D + B)^{-1}(b - T x_{k-1}).
\end{equation}
The matrix $D+B$ is lower triangular, and applying $(D + B)^{-1}$ would require either explicitly computing the inverse or performing a sequential forward substitution, both of which are computationally undesirable in many settings.
Thus, we approximate
\begin{equation}
(D + B)^{-1} \approx \left(\sum_{g=0}^{G} (-D^{-1}B)^g\right) D^{-1} \coloneqq \Omega D^{-1}
\end{equation}
using a truncated Neumann series.
Expressed in terms of the initial guess $x_0$, the approximation to the solution after $k$ Gauss-Seidel iterations is then given by
\begin{equation}\label{eq:TruncatedGaussSeidel}
\begin{split}
    x_k =& \sum_{j=0}^{k-1} (-1)^j \left(\Omega D^{-1} T\right)^j \Omega D^{-1} b  \\
    & + (-1)^k \left(\Omega D^{-1} T\right)^k x_0.
\end{split}
\end{equation}

\subsection{Quantum Gauss-Seidel via block encodings and LCU}

The truncated Gauss-Seidel scheme of Eq. \eqref{eq:TruncatedGaussSeidel} can be translated directly to a quantum setting using block encodings and LCU \cite{williams_quantum_2026}. In particular, the truncated Neumann series $\Omega$ and the outer summation over the iteration index are both realized by LCU procedures.
The qubit requirements therefore include $\log_2 N$ system qubits, $\lceil \log_2(G+1) \rceil$ ancillas for the truncated Neumann series, $\lceil \log_2(k+1) \rceil$ ancillas for the outer expansion, and additional qubits for the block-encoding multiplications, scaling linearly with $k$ and $G$.

\subsection{QSVT-based quantum approach}
As we see, the straightforward implementation of a quantum Gauss--Seidel scheme via block encodings and nested LCU constructions leads to a qubit overhead that scales linearly in both $k$ and $G$. Thus, we can expect to achieve a more resource-efficient implementation by using polynomial-based quantum techniques, as we have done for the quantum Jacobi algorithm.
However, QSVT requires the matrix it acts on to be normal, in order to result in a matrix polynomial that we are aiming for.
For the Gauss-Seidel scheme, we first examine the truncated Neumann series $\Omega$. The product $D^{-1}B$ is in general not normal, as it is the product of a diagonal matrix and a strictly lower triangular matrix. Therefore, it is not possible to generate the block encoding $U_\Omega$ with a QSVT subroutine. The same issue arises if we consider the classical pre-calculation of $\Omega D^{-1} T$ with the goal to perform a polynomial transformation on the resulting matrix. 
Thus it is not possible to use QSVT subroutines to implement the Gauss-Seidel method.

\bibliography{bibliography.bib}

\end{document}